\documentclass{sig-alternate-2013}

\usepackage{authblk}

\usepackage{subfig}
\def\etal{{\it et al.}}

\usepackage[hyphens]{url}

\usepackage{graphicx}
\usepackage{epstopdf}
\usepackage{color}
\usepackage{booktabs}

\newfont{\mycrnotice}{ptmr8t at 7pt}
\newfont{\myconfname}{ptmri8t at 7pt}
\let\crnotice\mycrnotice%
\let\confname\myconfname%

\permission{Permission to make digital or hard copies of all or part of this work for personal or classroom use is granted without fee provided that copies are not made or distributed for profit or commercial advantage and that copies bear this notice and the full citation on the first page. Copyrights for components of this work owned by others than ACM must be honored. Abstracting with credit is permitted. To copy otherwise, or republish, to post on servers or to redistribute to lists, requires prior specific permission and/or a fee. Request permissions from permissions@acm.org.}
\conferenceinfo{DH'15,}{May 18--20, 2015, Florence, Italy.\\
{\mycrnotice{Copyright is held by the owner/author(s). Publication rights licensed to ACM.}}}
\copyrightetc{ACM \the\acmcopyr}
\crdata{978-1-4503-3492-1/15/05\ ...\$15.00.\\
http://dx.doi.org/10.1145/2750511.2750524}

\begin{document}

\title{\#FoodPorn: Obesity Patterns in Culinary Interactions}

\author[1]{Yelena Mejova}
\author[1,2]{Hamed Haddadi}
\author[3]{Anastasios Noulas}
\author[1]{Ingmar Weber}

\affil[1]{Qatar Computing Research Institute, Qatar }
\affil[2]{Queen Mary University of London, UK }
\affil[3]{University of Cambridge, UK }

\maketitle
\begin{abstract}
We present a large-scale analysis of Instagram pictures taken at 164,753 restaurants by millions of users. Motivated by the obesity epidemic in the United States, our aim is three-fold: (i) to assess the relationship between fast food and chain restaurants and obesity, (ii) to better understand people's thoughts on and perceptions of their daily dining experiences, and (iii) to reveal the nature of social reinforcement and approval in the context of dietary health on social media. When we correlate the prominence of fast food restaurants in US counties with obesity, we find the Foursquare data to show a greater correlation at 0.424 than official survey data from the County Health Rankings would show. Our analysis further reveals a relationship between small businesses and local foods with better dietary health, with such restaurants getting more attention in areas of lower obesity. However, even in such areas, social approval favors the unhealthy foods high in sugar, with donut shops producing the most liked photos. Thus, the dietary landscape our study reveals is a complex ecosystem, with fast food playing a role alongside social interactions and personal perceptions, which often may be at odds.
\end{abstract}




\section{Introduction}
\label{sec:intro}

Food and dining is an important social and cultural experience, and today our social media feeds are filled with individuals checking in restaurants with friends, sharing both healthy and unhealthy dining experiences, and recommending restaurants and dishes to their social network\footnote{\url{http://www.psychologytoday.com/blog/comfort-cravings/201008/10-reasons-why-people-post-food-pictures-facebook}} -- to the point that some have suggested our obsession with food has grown to a ``food fetish''~\cite{FoodFetish}. But precisely because of its ubiquity and popularity, there are strong indicators for usefulness of these media~\cite{abbar2014you,conf/icwsm/PaulD11} to record the everyday interactions between individuals, society, and food.

Meanwhile, the alarming rise in economic and societal costs of obesity and diabetes have put these diet-related ailments on an ``epidemic'' scale~\cite{wang2007obesity}, and fast food has been widely cited as an important contributor~\cite{schlosser2012fast}. Recent studies have shown that neighborhoods with more fast food restaurants had significantly higher odds of diabetes and obesity~\cite{bodicoat2014number}, and, in particular, childhood obesity~\cite{currie2009effect}. These statistics, however, are static, and fail to capture the everyday dining outings of their participants, their thoughts and feelings about the food, and the social setting in which it takes place. Recently, the medical and healthcare communities have proposed utilizing social media data as a useful resource for monitoring people's eating habits and specifically those of obesity and diabetes patients~\cite{InnovativeHealth}. 

In this paper, we use data from two of the world's largest Online Social Networks (OSNs), namely Foursquare and Instagram, in order to study (\emph{i}) the relationship between fast food and chain restaurants to obesity, (\emph{ii}) the user thoughts, feelings, and perceptions of their dining experiences, and (\emph{iii}) social approval and interaction captured on these sites. Instagram is currently the world's most popular photo-sharing platform and with over 300 million users, it is bigger than Twitter. Using this rich source of data and social interactions, we focus on the pictures shared at 164,753 restaurants around the United States by over 3 million individuals. 

This data, we find, reveals more clearly the correlation between the prevalence of fast food restaurants in a county to its obesity rate, compared to statistics obtained by the Robert Wood Johnson Foundation and the University of Wisconsin Population Health Institute in 2013 County Health Rankings. Chain and fast food restaurants have fewer pictures and unique users visiting them, and pictures taken there receive both fewer likes and comments. Indeed, the connection between local restaurants and obesity is emphasized by the fact that the users from low-obesity regions use tags such as \texttt{\#smallbiz} and \texttt{\#eatlocal} much more frequently than in more obese areas. Even the users themselves realize the food in fast food places is unhealthy, with pictures taken there having twice as many tags associated with unhealthy topics. Unfortunately, social feedback (in terms of likes) often favors unhealthy restaurants -- donut, cupcake, and burger places -- despite individual users associating \texttt{\#foodporn} with predominantly Asian cuisines. These and other findings we describe here provide a window into the daily dining experiences of millions of people, potentially informing intervention and policy decisions.

The rest of this paper is organised as follows. In Section~\ref{sec:data} we describe the two datasets we use from Foursquare and Instagram. Sections~\ref{sec:resultsl} and~\ref{sec:perception} present the correlations between obesity rates and the prevalence of fast food or chain restaurants, and individuals' perception of their dining experiences, respectively. We then turn to the social approval and interaction prompted by pictures from various restaurants and with different tags in Section~\ref{sec:social}. We conclude with a brief overview of related work and further discussion our findings in its context in Sections~\ref{sec:related} and~\ref{sec:discussions}, respectively.

\section{Data collection}
\label{sec:data}

We begin by describing the two social media datasets we create: one of restaurant locations using Foursquare and another of picture posts using Instagram, and outline the US county-wide health and census data we use to supplement them.

\subsection{Foursquare locations}

\begin{figure*}[t!]
\subfloat[Population per county used\label{figure:population_county}]
  {\includegraphics[width=.30\linewidth]{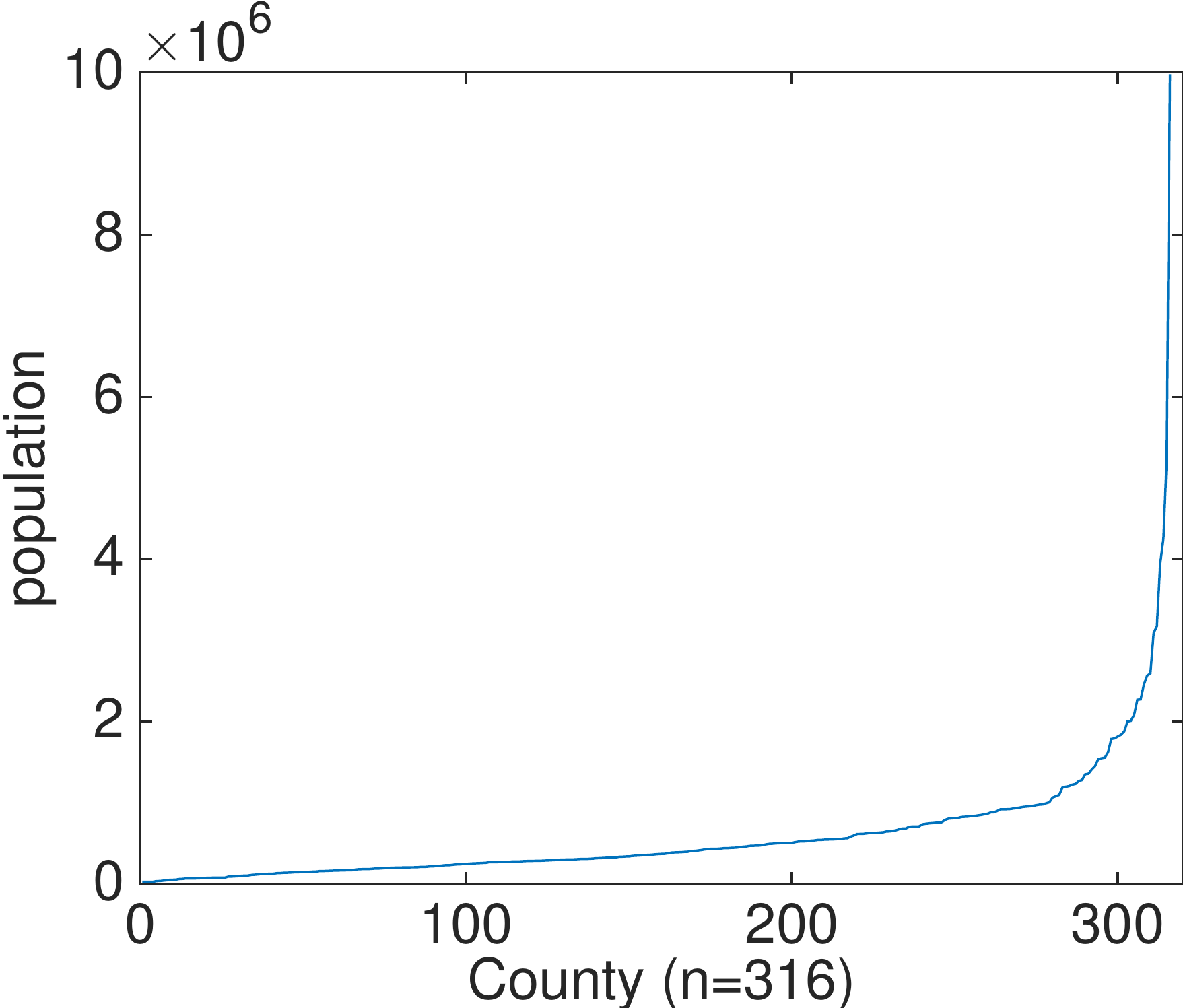}}\hfill
\subfloat[Pictures posted per county used\label{figure:countypics}]
  {\includegraphics[width=.29\linewidth]{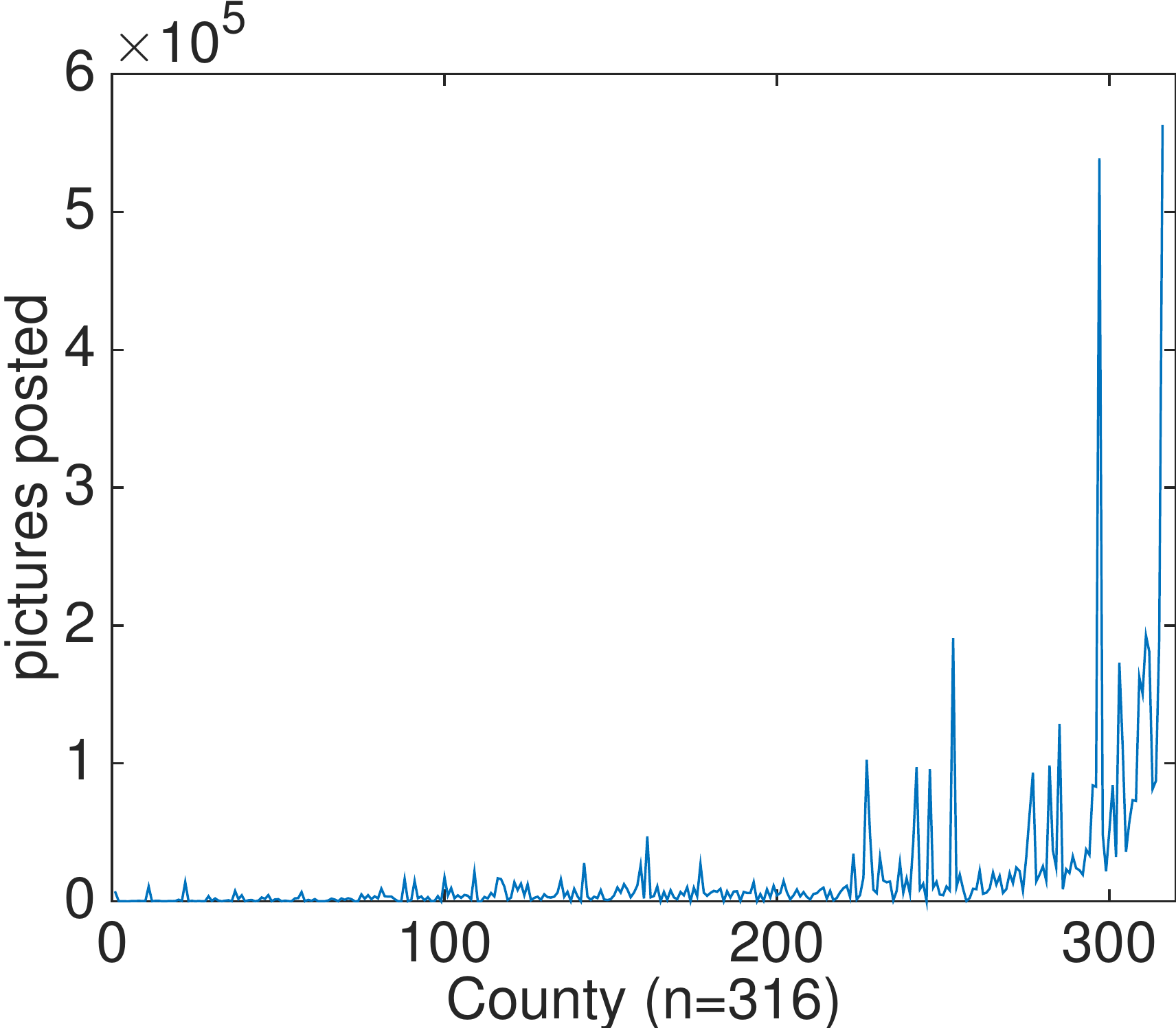}}\hfill
\subfloat[Unique users per county\label{figure:countyusers}]
  {\includegraphics[width=.30\linewidth]{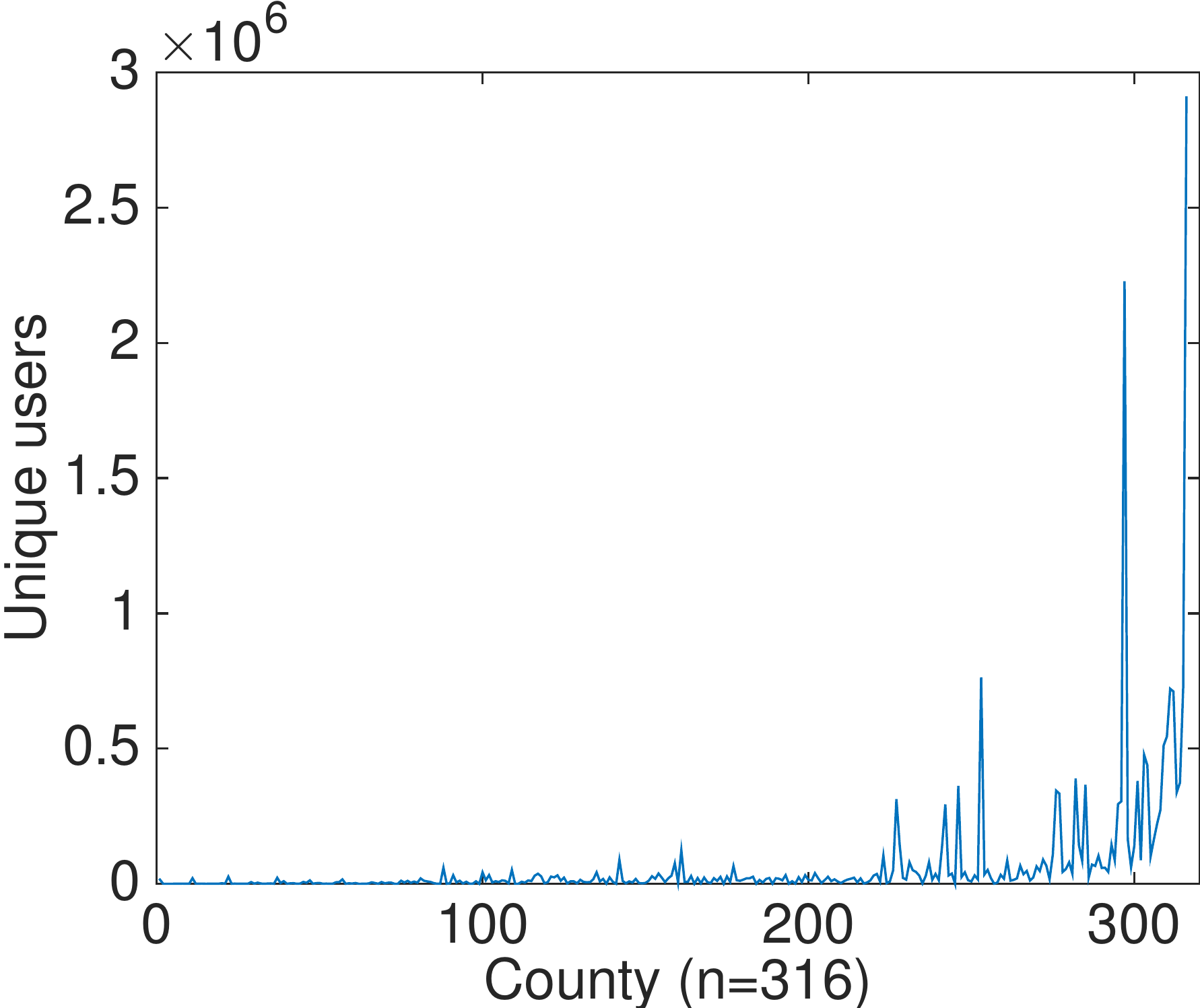}}
\caption{Populations, pictures posted, and unique users per US county in the Instagram dataset, ordered by population (shown in (a)) in all three graphs
\label{figure:datasetstats}}
\end{figure*}

\begin{figure}[t]
\centering
\includegraphics[width=0.8\linewidth]{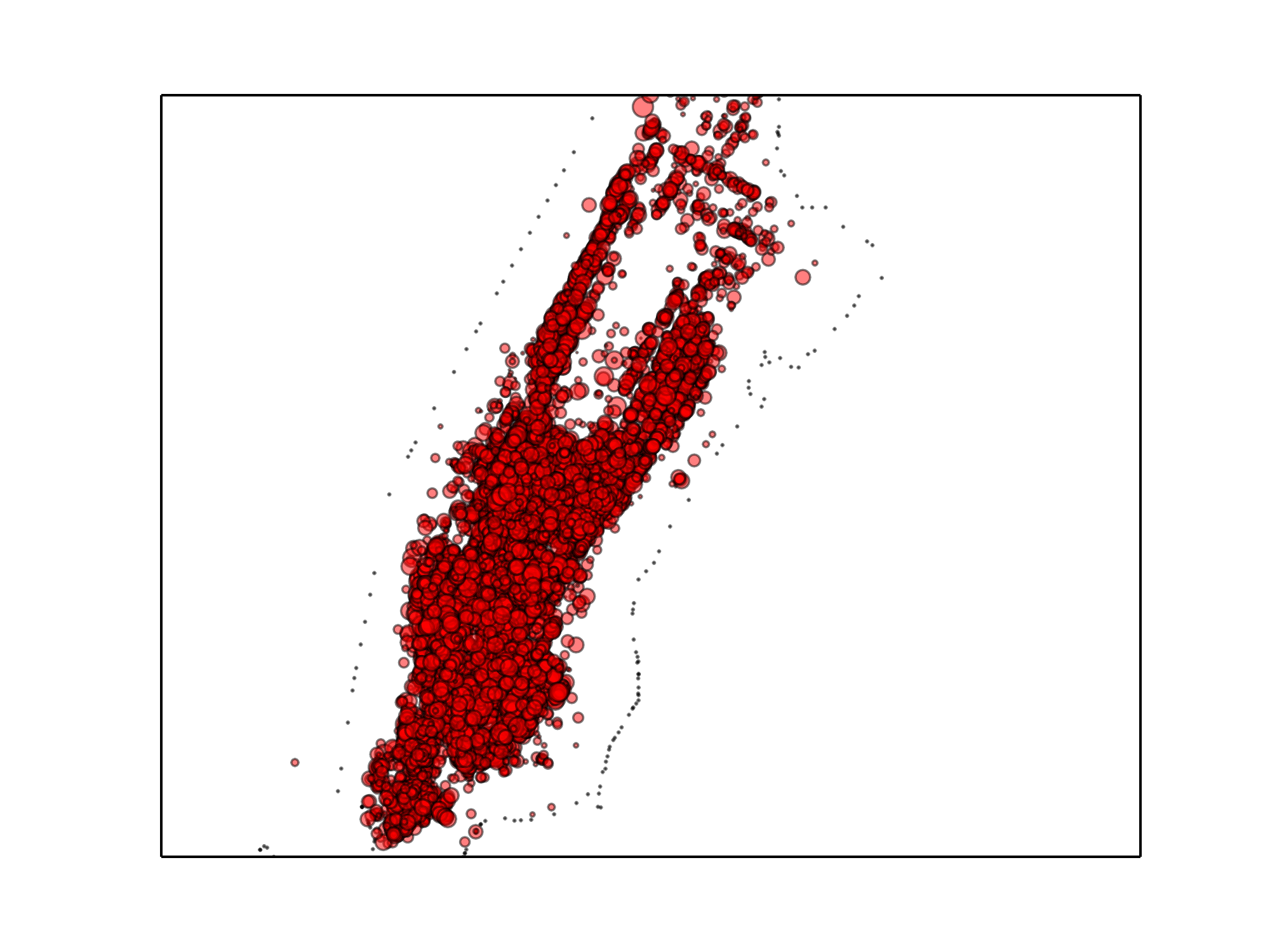}
\caption{Foursquare check-ins in Manhattan, New York}
\label{figure:4sq_nyc}
\end{figure}

Launched in 2009, Foursquare is a location-based service that users can exploit to discover places nearby and report their whereabouts to their online friends by ``checking in''. As of 2015, the service counts more than $55$ million users from around the world and numerous applications, including Instagram, have relied on Foursquare's API\footnote{\url{https://developer.foursquare.com}} to allow their users associate digital content to real world places. Here, we are using a dataset collected by Foursquare check-ins made public on Twitter during a 10-month time window (December 2010 - September 2011) that has yielded a set of 194,752 unique food places in the United States. To maintain a high level of data quality, Foursquare has been applying filters for the detection of fake check-ins or locations, in addition to putting in place a \textit{venue harmonization} process that removes duplicate entries in its venue dataset. Given the geographic coordinates of a Foursquare place, we have associated it with the county and state it belongs to using the Data Science Toolkit API\footnote{\url{www.datasciencetoolkit.org}}. Figure \ref{figure:4sq_nyc} shows check-ins in the Manhattan Island of New York City, illustrating the dense coverage this data provides.

\subsection{Instagram pictures}

Next, we map Foursquare locations to the Instagram ones using the Instagram Location Endpoints API\footnote{\url{http://instagram.com/developer/endpoints/locations/}} on September 2014. This lead to 164K unique Instagram locations, as it is possible that a Foursquare location does not map to a corresponding Instagram location ID. At the time of the collection, Foursquare was the location provider for Instagram queries, hence this conversion is highly accurate for our specific purpose. 

\begin{figure}[t]
\centering
\includegraphics[width=\linewidth]{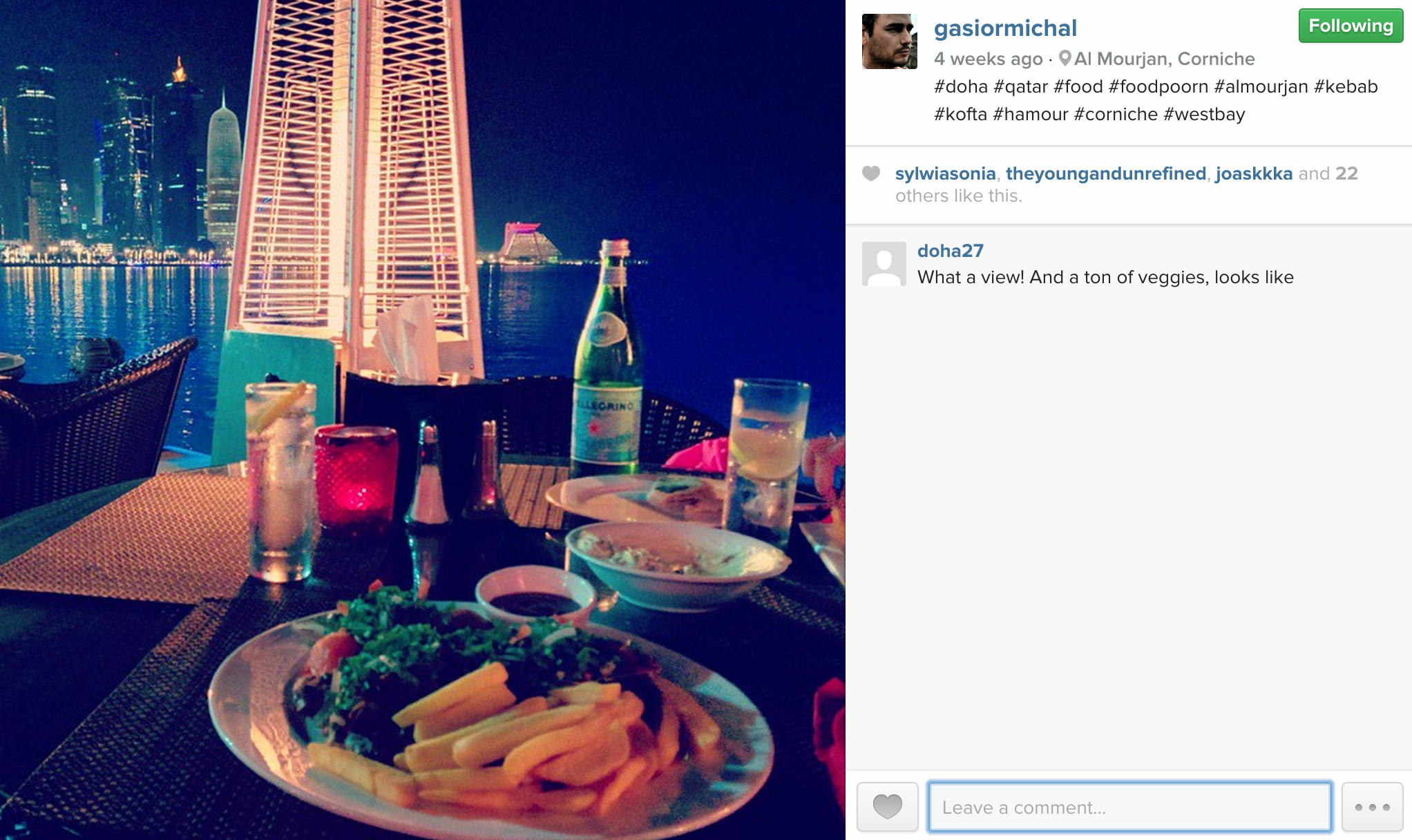}
\caption{Example of an Instagram post, associated with a restaurant, with hashtags in description, likes, and a comment}
\label{figure:screenshot}
\end{figure}

We then query each location using the location endpoints API to get a list of the recent posted media. Users can only post media to locations which are within a maximum of 5,000 meters radius and they have to choose a location from within the Instagram app, hence the mapping to locations is reliable. This query returns a list of recent images from each given location, with their corresponding hashtags, comments, likes, and ID of the users involved (such as one illustrated in Figure~\ref{figure:screenshot}). Because adding location is turned off by default in the app, the pictures collected are voluntarily and purposefully tagged with location by the users. The recent image posts were collected once per month, throughout September, October, and November 2014. This lead to retrieving information on 20,848,190 Instagram posts from 3,367,777 unique individuals from 316 US counties. It is important to notice that, like most other OSNs, our Instagram data has inherent biases, as it naturally represents only a small percentage of the US population from states which have higher activity levels on social media. Yet it is a very rich source of data, including context (through tags) and social interaction (through comments and likes). In preparing the data, we took the conservative approach of removing all the duplicate images (by URL and owner) which may have been re-shared by other users through third party apps. This step ensured we are only dealing with original, unique posts. 

\vspace{0.5cm}

\subsection{County health \& census data}

We use the 2013 County Health Rankings\footnote{\url{http://www.countyhealthrankings.org/about-project}} (CHR), which is a collaboration between the Robert Wood Johnson Foundation and the University of Wisconsin Population Health Institute. It provides vital health factors including obesity and diabetes rates, demographics including income and education, and community variables including prevalence of fast food restaurants, in each county in America. In particular, we focus on the obesity rate, which ranges from counties with 13\% obesity (Teton, Wyoming) to 48\% (Greene, Alabama), as well as the percentage of fast food restaurants.

\subsection{Fast food restaurants}

Among its meta-data, the Foursquare dataset contains a classification of restaurants, which includes fast food. However, there is often variability in the way individual restaurants are associated with a class, with some fast food chains sometimes having Burger designation. In attempt to at least capture the most prominent fast food chains, we supplement this knowledge with the list of top 50 fast food chains by sales\footnote{\url{http://www.qsrmagazine.com/reports/qsr50-2012-top-50-chart}} and Wikipedia lists of fast-food and fast-casual chains in the US\footnote{\url{https://en.wikipedia.org/wiki/List_of_restaurant_chains_in_the_United_States}}. We then consider all restaurants in the Fast Food and Burger categories of Foursquare or in these lists to be a fast food place.

\subsection{Population representativeness}

Finally, we check whether the number of users our sample contains is representative of the overall population of the counties, summarized in Figures~\ref{figure:datasetstats}. We find that Spearman's rank correlation $\rho$ between the county population and the unique users is 0.746. We also find the users to contribute proportional number of pictures, with $\rho$=0.9954 between number of users and that of pictures per county. Thus, although the subset is likely to be of tech-savvy and young people, it is at least proportional to the overall population.

\vspace{1cm}
\section{Fast Food}
\label{sec:resultsl}

In his book, ``Fast Food Nation: The Dark Side of the All-American Meal'', Eric Schlosser puts fast food as a central player in the rise of obesity in US \cite{schlosser2012fast}. Here, we examine to what extent data obtained from CHR and that obtained using social media reveal the relationship between the prominence of fast food restaurants and obesity.

\subsection{Fast food and obesity}

\begin{figure*}[t!]
\subfloat[CHR\label{figure:obesity_vs_pff_chr}]
  {\includegraphics[width=.33\linewidth]{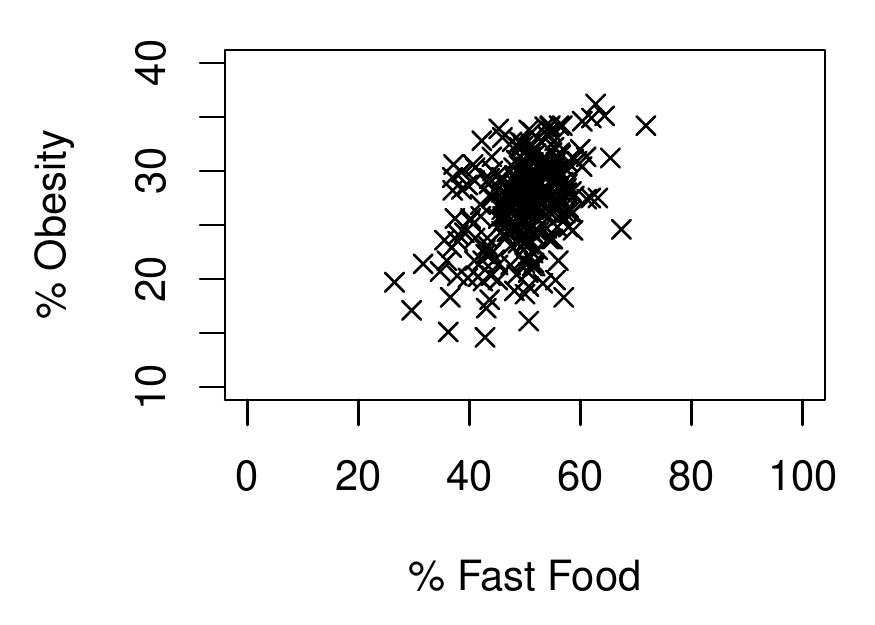}}\hfill
\subfloat[Foursquare\label{figure:obesity_vs_pff_fsq}]
  {\includegraphics[width=.33\linewidth]{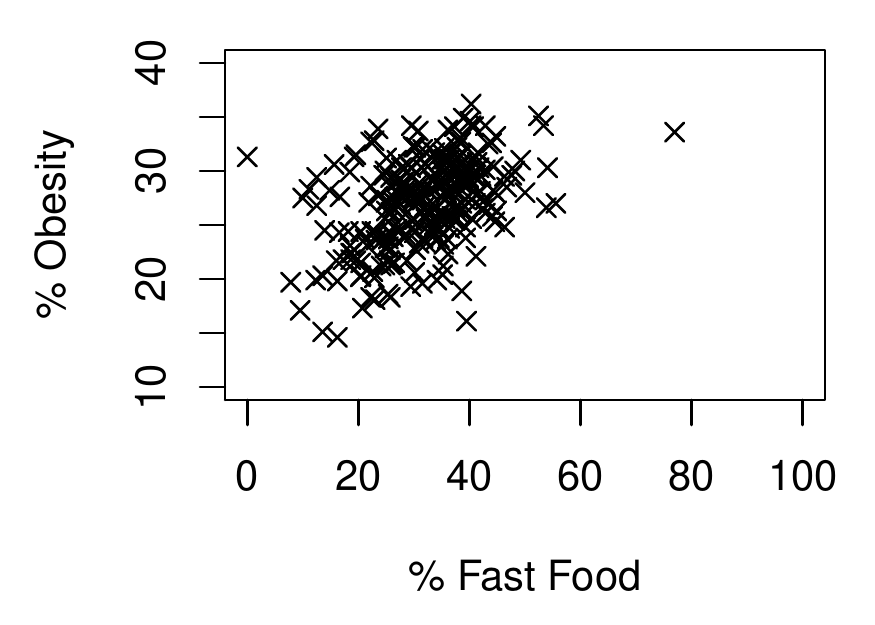}}\hfill
\subfloat[Local/Chain\label{figure:obesity_vs_local}]
  {\includegraphics[width=.33\linewidth]{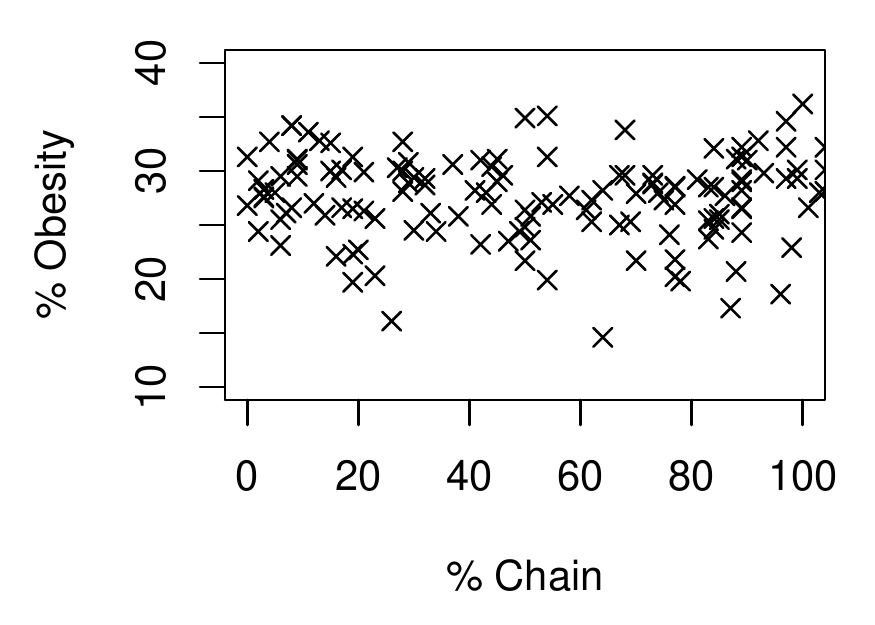}}\hfill
\caption{County-wide percentage of fast food places as measured by CHR (a) or Foursquare (b), and number of chain restaurants (c) to percentage of obesity}
\label{figure:obesity_vs_pff}
\end{figure*}

Are fast food restaurants correlated with obesity? Here, we consider only the counties which have more than 5 unique restaurants in our Foursquare dataset ($N$=280). We first examine the County Health Rankings (CHR) data, correlating percentage of fast food places to percentage of obese population, with a resulting Pearson correlation of 0.246. When we weight the means by population (using weighted correlation), we get a slightly higher one, of 0.267. Now, an alternative source of information is the share of fast food places is our Foursquare dataset, in which the restaurants are present only if a check-in took place. We find a substantially higher correlation of 0.424. The relationship between obesity rate with percent fast food places for the two sources of data is shown in Figures \ref{figure:obesity_vs_pff}a,b. 

Note that our data shows a much less prominent share of fast food restaurants. This may be due to our definition of fast food restaurants (see previous section) differing from that of CHR (unfortunately, their definition is not publicly available). Alternatively, social media users may label the restaurants as something other than fast food. Finally, they may not check in to fast food restaurants as much, and we are capturing the peculiar dining behavior of the population active on Foursquare. 

We also check the correlation of the prevalence of fast food restaurants and various demographics (available in the CHR data). Among race, income, and poverty-related variables, the highest was the correlation with the population of children under 18 years of age at 0.499 for CHR and 0.450 for Foursquare data, indicating higher exposure of families and kids to these restaurants.

\subsection{Local places versus chain restaurants}

\begin{figure}[h]
\vspace{0.2cm}
\begin{centering}
\subfloat[\label{figure:fast_picsperrest}]
  {\includegraphics[width=.235\linewidth]{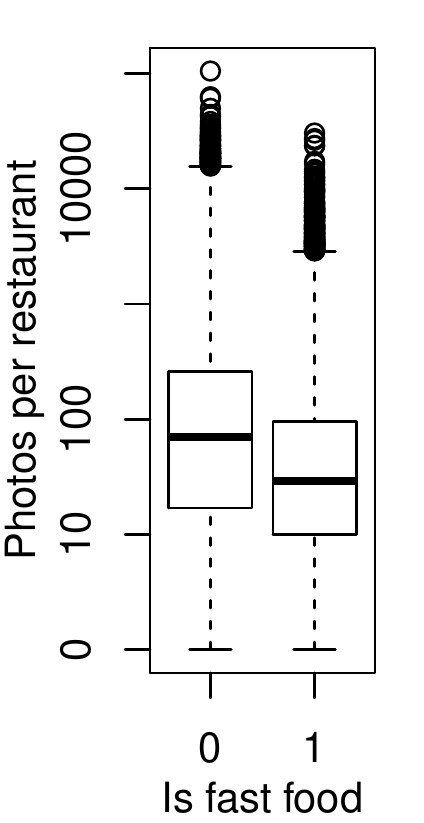}}\hspace{0.05cm}
\subfloat[\label{figure:fast_picsperuser}]
  {\includegraphics[width=.235\linewidth]{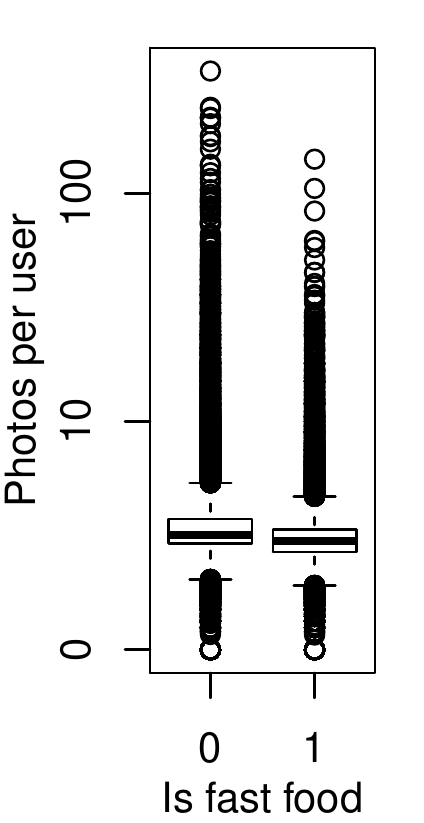}}\hspace{0.05cm}
\subfloat[\label{figure:local_picsperrest}]
  {\includegraphics[width=.235\linewidth]{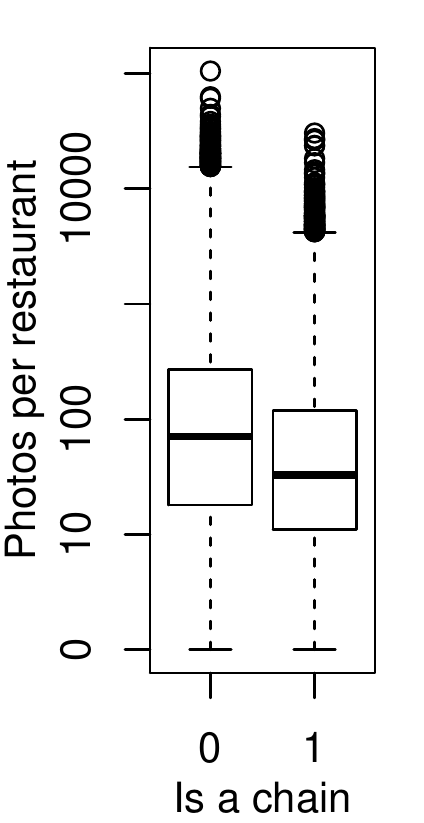}}\hspace{0.05cm}
\subfloat[\label{figure:local_picsperuser}]
  {\includegraphics[width=.235\linewidth]{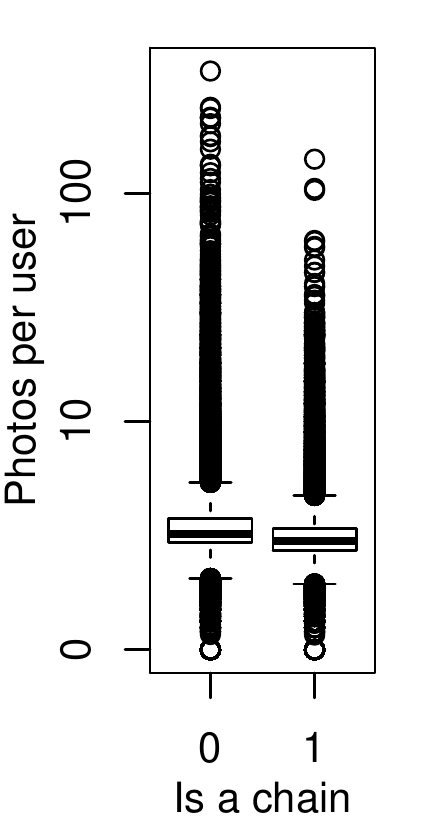}}
\caption{Distribution of per-restaurant statistics: number of photos posted for a restaurant (a,c) and average number of photos each user posted (b,d) for fast-food vs slow-food (a,b) and local vs chain (c,d) restaurants (y-axis is log-scale)}
\label{figure:localfast_pics}
\vspace{0cm}
\end{centering}\end{figure}

Because our dataset allows us to get an aggregate view of all restaurants Foursquare users visit, we also take a data-driven approach at defining what is a chain restaurant. Concretely, across the 164,753 unique Foursquare locations we looked for exact repetitions in place names, irrespective of upper or lower case. The most frequent repetition was ``Starbucks'' (4,870) followed by other popular chains. Any name repeated 10 or more times was then considered a chain. Other places exactly matching, up to casing, or containing such a location, such as ``Starbucks at Super Target'' were then considered as part of a chain. This worked well and also caught frequent variants such as both ``McDonald's'' and ``McDonalds''. Manual inspection showed that the minimum threshold of 10 worked well to distinguish between proper chains and merely frequently repeated restaurant names such as ``Asia Wok''. The threshold was also low enough to allow a restaurant to have a handful of local branches, without being considered a chain. Out of the chain restaurants we detected, 30\% were not designated as fast food (i.e. they are ``slow'' food), suggesting that most chains are fast food, but not all (whereas only 8\% of fast food restaurants were not detected as chain). Since 72\% of our dataset is slow food, chains are overwhelmingly likely to be fast food, compared to the overall distribution.

Figure \ref{figure:obesity_vs_local} shows the relation of the number of chain restaurants in a county to \% obesity. There is no clear correlation between obesity and these restaurants. However, we do find a large discrepancy in the activity of Instagram users -- they are much more likely to share a photo from a local restaurant than a chain. On average, there are 150 photos shared from a chain restaurant, compared to 396 at a local one.

We also find that although chain restaurants benefit from their name recognizability, small local restaurants have more dedicated clientele. Indeed, on average chain restaurants have 46 distinct users posting at least one picture, whereas local chains get 107 users. Per-user, there is also a difference, with those going to local restaurants sharing 3.6 images compared to 3.2 at a chain. Similar observations can be made for fast vs slow food. Figures \ref{figure:localfast_pics}(a-d) illustrate these statistics, with long tails of extremely active restaurants (note that y-axis is in log scale) for local and slow food restaurants.

\section{Food Perception}
\label{sec:perception}

Users tag their pictures to provide the context in which the dietary experience took place. First, we turn our attention to the most used hashtag in our dataset, which is also one unambiguously expressing delight in a dietary experience: \texttt{\#foodporn}. We track this hashtag across the restaurant categories, and the resulting top and bottom restaurants by shares of the tag are listed in Table \ref{tbl:categoryfoodporn}. Asian foods, including Indonesian, Malaysian, and Vietnamese, dominate the top, along with Molecular Gastronomy, which uses technical innovations to create new textures and dishes. At the bottom we see drinking establishments (which is understandable, since \texttt{\#foodporn} is about food, not drinks), as well as Swiss, German, and Fast food. Also, Wings and Fish \& Chips join Gluten-free as the least exciting restaurants. This ranking is in a stark difference to the number of pictures associated with the categories (shown in $N_{pic}$ column), with American, Coffee Shop, and Mexican being the most popular, suggesting that the food people love often does not come from their everyday visitations. But it is not necessarily the case that exciting foods come from little-posted places. In fact, the correlation between the \texttt{\#foodporn} hashtag and number of pictures is only slightly negative at -0.174.

\begin{table}[h]
\caption{Top and bottom restaurant categories by shares of tag \#foodporn, along with their overall number of pictures
\label{tbl:categoryfoodporn}}
\begin{center}
\footnotesize
\begin{tabular}{llrr}
\toprule
 & \textbf{Category} & \textbf{\#FP} & \textbf{$N_{pic}$} \\
\midrule
1 & Malaysian & 0.21 & 6,169 \\
2 & Vietnamese & 0.17 & 119,915 \\
3 & Indonesian & 0.17 & 257 \\
4 & Dumplings & 0.17 & 8,092 \\
5 & Molecular Gastronomy & 0.16 & 6,668 \\
6 & Korean & 0.16 & 115,898 \\
7 & Peruvian & 0.15 & 7,472 \\
8 & Mac \& Cheese & 0.15 & 6,154 \\
9 & Thai & 0.14 & 119,563 \\
10 & Japanese & 0.14 & 260,188 \\
11 & Dim Sum & 0.14 & 24,322 \\
12 & Filipino & 0.13 & 6,138 \\
13 & Sushi & 0.12 & 425,425 \\
14 & Chinese & 0.12 & 178,512 \\
15 & Asian & 0.12 & 210,087 \\
\midrule
66 & Fish \& Chips & 0.06 & 538 \\
67 & Gluten-free & 0.06 & 614 \\
68 & Gastropub & 0.05 & 139,059 \\
69 & Fast Food & 0.05 & 114,103 \\
70 & Tea Room & 0.05 & 69,477 \\
71 & Food & 0.05 & 21,619 \\
72 & Wings & 0.05 & 98,737 \\
73 & Swiss & 0.05 & 4,198 \\
74 & German & 0.04 & 38,412 \\
75 & Juice Bar & 0.04 & 38,944 \\
76 & Brewery & 0.02 & 284,258 \\
77 & Afghan & 0.02 & 229 \\
78 & Coffee Shop & 0.02 & 848,212 \\
79 & Winery & 0.01 & 31,229 \\
80 & Distillery & 0.00 & 2,315 \\
\bottomrule
\end{tabular}
\end{center}
\end{table}

Note that whereas \texttt{\#foodporn} is the most frequently used hashtag (1,890,691 occurrences, followed by \texttt{\#food}, \texttt{\#nyc}, \texttt{\#yum}, \texttt{\#love}, \texttt{\#dinner}, and  \texttt{\#instagood}), among the frequent hashtags, one associated with a camera application \texttt{\#vscocam}\footnote{\url{https://play.google.com/store/apps/details?id=com.vsco.cam&hl=en}} is the most liked one (on average and in median). However, \texttt{\#instahealth}, which is associated with health and motivation, is even more liked when used, even though it is not in the top 50.

\subsection{Hashtag labeling}

Beyond \texttt{\#foodporn}, among the many dimensions present in these tags, we focus on those concerning \emph{health}, \emph{emotion}, and \emph{social} aspects. We use CrowdFlower\footnote{\url{http://www.crowdflower.com/}} platform to crowdsource the labeling of the top 2,000 used tags. The labeling of each dimension involved a worker to complete a task consisting of 10 words. For each word, we required 3 labels by different annotators. The experiments ran in ``quiz'' mode, requiring correct answers for 3 out of 4 posed gold-standard questions to begin the task, and with 1 test question in every following task for quality control. The agreement was high, with label overlap at 92-99\%. Our labeling effort resulted in four binary variables: \emph{healthy} (refers to healthy food), \emph{unhealthy} (refers to unhealthy food), \emph{social} (refers to a social setting), and \emph{emotion} (expresses some emotion). We made this term list available to download online\footnote{\url{http://bit.ly/14ROSvn}}.

\subsection{Perception of fast food}

\begin{figure}[t]
\begin{center}
 \includegraphics[scale=0.58]{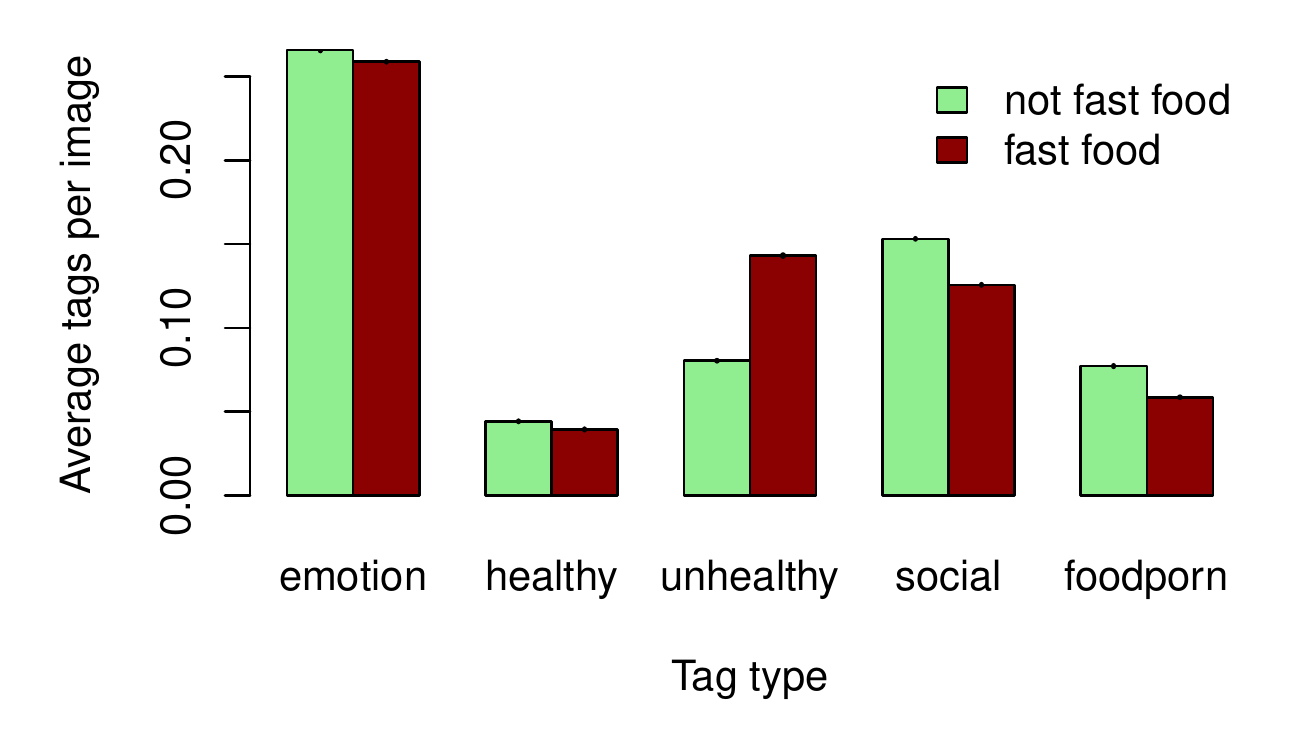}\end{center} 
\caption{Proportion of pictures with tags of a type (and single hashtag \#foodporn) in fast food versus all other restaurants}
\label{figure:ff_tagtype}
\end{figure}

Using the above hashtag classifications, we now can observe the extent to which each tag group was used in pictures taken at fast food restaurants versus all others. Figure \ref{figure:ff_tagtype} shows the proportion of the use of these tags. Hashtags that have to do with emotion are the most used, and about healthy foods the least. We see a marked difference between the use of unhealthy food tags, with fast food showing twice as much use as slow food, indicating that the users perceive it to be unhealthy. However, when we compare the use of \texttt{\#foodporn}, we see it used more for slow food restaurants. Social occasions are also more present in slow food places. Keep in mind that the sensitivity of this method depends on the extent of the vocabulary, and the particular percentages of the images are not as informative as their comparative proportion.

When we look at chains versus local, we find that, on the contrary, users tended to post hashtags related to social events at chain restaurants (at 0.158) more than local (0.137) (significant at $p$<0.001), suggesting that chain restaurants which may not be labeled as fast food (such as \emph{The Cheesecake Factory} or \emph{California Pizza Kitchen}) play an important role in people's social outings.

\subsection{Chatter around obesity}

Beyond the fast food restaurants, we are interested in the user activities and food perception that can be associated with areas of high obesity. Thus, we segment the data into low, middle, and high terciles by the percentage of obese residents, with the breaks at 21.8\% and 29.0\%, which were computed in the range of the counties in our dataset. Whereas the top ranking hashtags in each tercile are quite similar (dominated by \texttt{\#foodporn}, \texttt{\#food}, and \texttt{\#yum}), when we subtract the rankings of lowest from highest tercile, we reveal tags which are associated more with high or low obesity regions.

\begin{table}[t]
\caption{Health and perception-related hashtags more prominent in high and low obesity counties, determined by prominence rank difference}
\begin{center}
\footnotesize
\begin{tabular}{lll}
\toprule
\textbf{List} & \textbf{High Obesity} & \textbf{Low Obesity} \\
\midrule
100 & \texttt{\#foodstyling} & \texttt{\#smallbiz} \\ 
    & \texttt{\#ilovesharingfood} & \texttt{\#guiltfree} \\ 
    & \texttt{\#foodstamping} & \texttt{\#whodat} \\ 
    & \texttt{\#foodphoto} & \texttt{\#mycurrentsituation} \\ 
    & \texttt{\#fatlife} &  \\ 
    & \texttt{\#f52grams} &  \\ 
\midrule
500 & \texttt{\#tacotuesday} & \texttt{\#myview} \\ 
    & \texttt{\#foodforfoodies} & \texttt{\#eatlocal} \\ 
    & \texttt{\#firsttime} & \texttt{\#plantbased} \\ 
    & \texttt{\#finedining} & \texttt{\#smoke} \\ 
    & \texttt{\#sharefood} & \texttt{\#eatwell} \\
\bottomrule
\end{tabular}
\label{tbl:tags_highlowobesity}
\end{center}
\end{table}

Table~\ref{tbl:tags_highlowobesity} shows a selection of hashtags which have the most different rankings between the lowest and highest terciles. Concretely, we considered top 20 terms in two lists that were filtered according to the number of minimum occurrence counts requited ($>=100$ for more strict or $>=500$ for more inclusive lists). Note that this is not a complete list of terms, most of which are places, cuisines, and particular restaurants, thus the tags in the table are those left after excluding the above. First, we notice food sharing tags, including \texttt{\#sharefood} and \texttt{\#ilovesharingfood}, to appear in the high obesity list, as well as \texttt{\#fatlife}. Interestingly, whereas \texttt{\#desserts} is more prominent in high obese areas, the singular term \texttt{\#dessert} is the 17th most popular tag in low obesity areas -- showing a fine distinction between having one versus more than one dessert. On the low obesity list we find tags associated with healthy food -- \texttt{\#guiltfree}, \texttt{\#eatwell}, \texttt{\#plantbased} -- and those referring to local foods -- \texttt{\#smallbiz}, \texttt{\#eatlocal}. There were also more individualistic tags such as \texttt{\#mycurrentsituation} and \texttt{\#myview}. Once again, we see an association between small local restaurants and healthier communities.

\section{Social approval}
\label{sec:social}

\begin{figure}[t!]
\subfloat[Likes\label{figure:obesity_likes}]
  {\includegraphics[width=.45\linewidth]{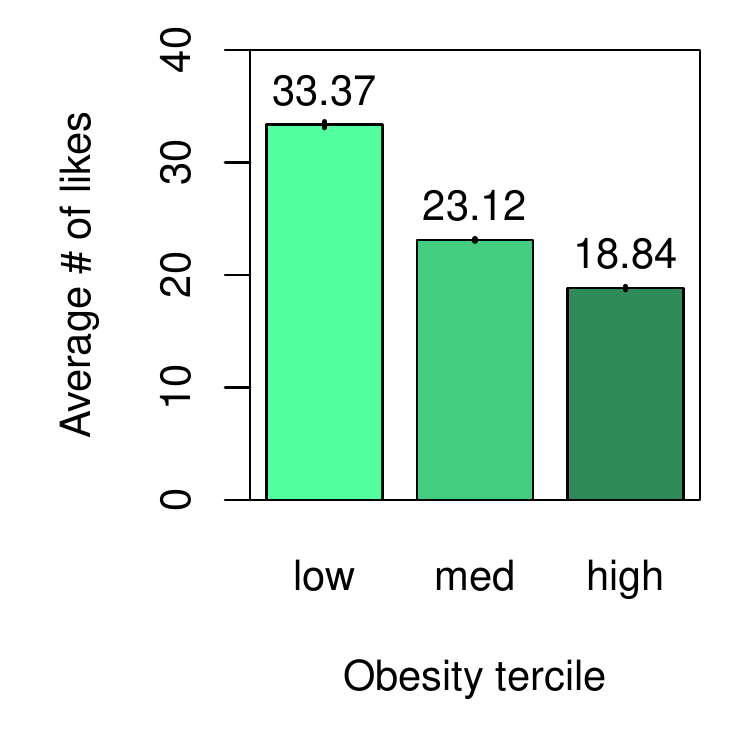}}\hfill
\subfloat[Comments\label{figure:obesity_comm}]
  {\includegraphics[width=.45\linewidth]{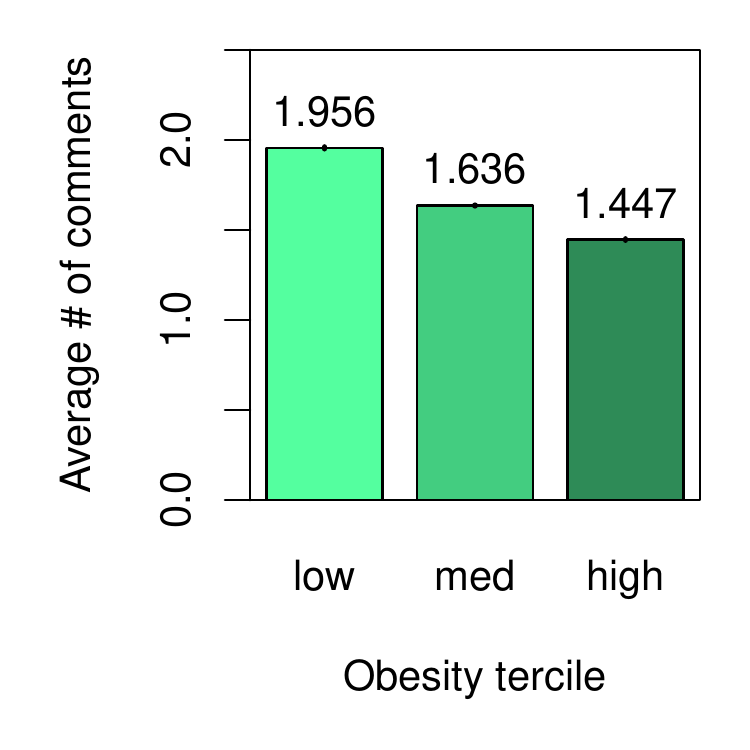}}\hfill
\caption{Average likes and comments for pictures taken in places with low, medium, and high obesity rates}
\label{figure:obesity_likes_comm}
\end{figure}

Finally, we turn to variables we associate with social interaction, and, more concretely, approval and engagement -- likes and comments the picture receives. Figure \ref{figure:obesity_likes_comm} shows the average number of likes (a) and comments (b) for a picture from areas with low, medium, and high obesity rate (as defined in the previous section). We find a large distinction between the implicit approval in term of likes and conversation engagement in terms of comments between the three groups. Pictures taken in high obesity areas tend to have fewer likes and comments, and in terms of likes, 56\% fewer than their counterparts coming from low obesity areas.

\begin{figure}[t!]
\centering
\includegraphics[width=0.9\linewidth]{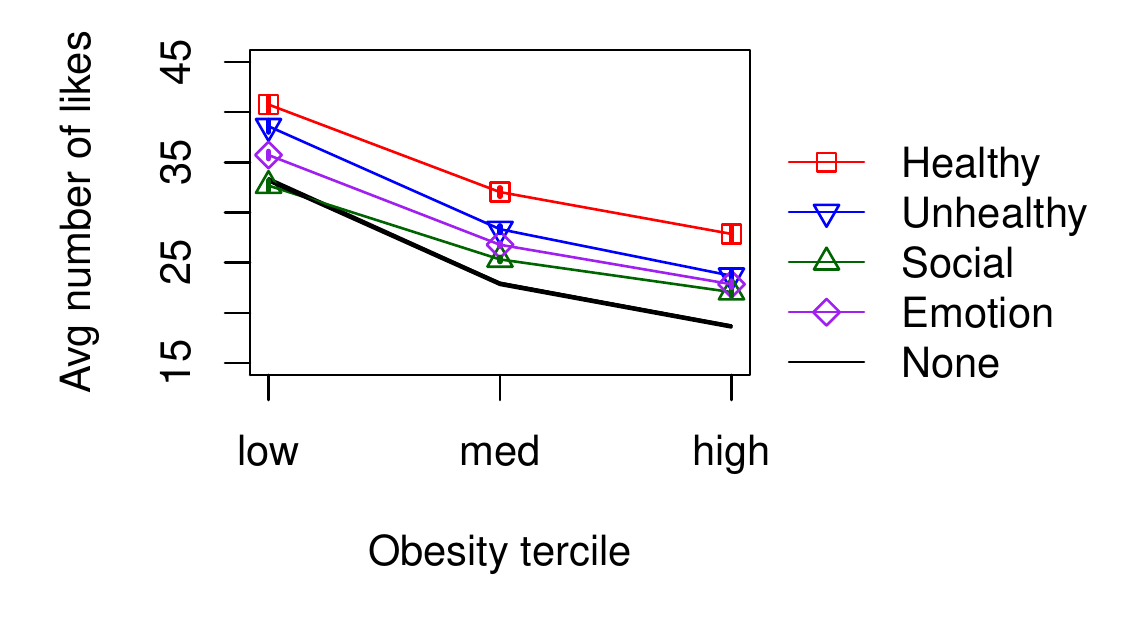}  
\caption{Average number of likes for pictures having or not having a certain type of hashtag, with 95\% confidence bars}
\label{figure:clike_ttypes}
\end{figure}

Considering the tags associated with the pictures, we also find a distinct difference between the likes and comments when certain types of tags are present. Figure \ref{figure:clike_ttypes} shows the average number of likes for pictures with certain tags (for pictures having at least one tag). We observe that both healthy and unhealthy tags provoke more likes, echoing the observation of~\cite{Adolescents} that there are societal pressures both for and against unhealthy lifestyles. In low obesity areas healthy and unhealthy tags are associated with (roughly) a similar number of likes whereas this difference is more pronounced in high obesity areas. Interestingly, ``social'' gains in popularity compared to ``none''. Is being gregarious linked with obesity? Further study of individuals' behavior would shed light on this finding. Similar observations can be made for the number of comments (however, we find that there are also substantially more comments for social tags), and we omit the plots in the interest of space.

Notably, we found rather small distinctions between both of these variables for local/chain or fast/slow food restaurants, suggesting that the restaurants themselves affect the social engagement much less than the user's presentation of the experience in a form of hashtags. Indeed, the hashtag \texttt{\#foodporn} prompts on average 38 likes, compared to pictures without it at 26 likes (similarly with comments at 2.5 with and 1.7 without).

Finally, we find that pictures of dessert are most liked in our dataset, followed by mac \& cheese, burgers, and French food (see Table~\ref{tbl:categorylikes}). Interestingly, donuts and cupcakes top the rankings for low-obesity areas, and instead New American and street food dominate in high-obesity ones (full lists omitted due to space). Overall, the foods at the top of this list are high in sugar, fat, and carbohydrates, which have been shown to have addictive properties~\cite{drewnowski1992taste,ventura2014neurobiologic}. Furthermore, the overwhelming approval of donuts in the low-obesity tercile is surprising, at 70 likes on average per picture, showing that the online interactions may bring out our desires, but not necessarily illustrate the typical offline behavior.

\begin{table}[t]
\caption{Top and bottom restaurant categories by average number of likes per picture, and overall number of pictures
\label{tbl:categorylikes}}
\begin{center}
\footnotesize
\begin{tabular}{llrr}
\toprule
 & \textbf{Category} & \textbf{$\mu_{likes}$} & \textbf{$N_{pic}$} \\
\midrule
1 & Donuts & 49.1 & 97,370 \\
2 & Cupcakes & 45.9 & 72,685 \\
3 & Juice Bar & 38.1 & 38,944 \\
4 & Mac \& Cheese & 37.9 & 6,154 \\
5 & Burgers & 37.8 & 429,170 \\
6 & French & 37.3 & 215,522 \\
7 & Desserts & 37.0 & 185,141 \\
8 & Italian & 37.0 & 502,339 \\
9 & Filipino & 36.5 & 6,138 \\
10 & Mediterranean & 36.1 & 74,809 \\
11 & Ice Cream & 36.1 & 226,659 \\
12 & Bakery & 35.4 & 245,074 \\
13 & Brazilian & 35.3 & 37,646 \\
14 & Japanese & 34.8 & 260,188 \\
15 & Bagels & 34.0 & 26,900 \\
\midrule
66 & Indian & 23.5 & 43,142 \\
67 & Tapas & 23.5 & 74,960 \\
68 & Moroccan & 23.4 & 4,085 \\
69 & Paella & 23.2 & 2,874 \\
70 & African & 22.4 & 6,372 \\
71 & Scandinavian & 22.3 & 3,748 \\
72 & Portuguese & 22.2 & 1,103 \\
73 & Arepas & 21.7 & 2,715 \\
74 & Molecular Gastronomy & 20.9 & 6,668 \\
75 & Fish \& Chips & 20.8 & 538 \\
76 & Gluten-free & 20.7 & 614 \\
77 & Indonesian & 20.5 & 257 \\
78 & Ethiopian & 18.8 & 7,536 \\
79 & Distillery & 18.0 & 2,315 \\
80 & Mongolian & 14.8 & 1,044 \\
\bottomrule
\end{tabular}
\end{center}
\end{table}

\section{Related research}
\label{sec:related}


Our study contributes to a growing body of work which uses social media for monitoring health-related activity. The potential for such studies to provide insights into larger scale societal behaviour trends and social interactions presents unique advantage over standard methods of tracking dietary behavior, such as food diaries. 
Recently, Culotta~\cite{Culotta:2014:ECH:2556288.2557139} performed a linguistic analysis of tweets from US users and found many categories that were significant predictors of health statistics including teen pregnancy, health insurance coverage, and obesity. 
Silva~\etal~\cite{Silva14:youare} use 5 million Foursquare check-ins (using Twitter data) and survey data, finding strong temporal and spatial correlation between individuals' cultural preferences and their eating and drinking habits. 
Fried~\etal~\cite{DBLP:journals/corr/FriedSKHB14} also predict overweight and diabetes rates for 15 largest US cities using language-based models built on three million food-related tweets, achieving an accuracy of 80.39\% on the task of predicting whether the overweight rate is below or above the median. In this paper, we have use a large dataset designed to capture the dining out patterns of social media users across the United States. Unlike previous endeavors, we focus specifically on fast food and chain restaurants, as well as other restaurant categories, as they relate to obesity and food perception.


Fast food has been widely cited as a contributor to obesity and related ailments \cite{schlosser2012fast}. Studies have shown, for example, that neighborhoods (of 500m radius) with more fast food restaurants had significantly higher odds of diabetes and obesity, with one additional diabetes case for every two new restaurants, assuming causal relationship \cite{bodicoat2014number}. The effect is especially pronounced for children, with those studying within a tenth of mile of a fast food restaurant having a 5.2\% higher chance to be obese \cite{currie2009effect} (however, these effects are not witnessed at larger distances). Also, using County Health Rankings, as well as US Department of Agriculture (USDA) and CDC Food Atlas, Newman~\etal~\cite{newman2014implications} have recently concluded that ``higher levels of fast food restaurant saturation are associated with increased levels of childhood obesity in both urban and poor areas, with the largest negative effect of fast food availability on obesity occurring in more economically disadvantaged, urban areas''. Our social media data confirms the link between obesity and the density of fast food restaurants, even amplifying the effect beyond the available statistics.


Besides the quantitative statistics, social media provides a lexicon of hashtags to put the dining experience in personal and social context. When using Twitter to build language models for distinguishing cities above and below median overweight percentage, Fried~\etal~\cite{DBLP:journals/corr/FriedSKHB14} find the most distinguishing features to be pronouns, as well as particular kitchens. Due to the social nature of Instagram, we further find that the food sharing behavior is higher for high-obesity areas, whereas mentioning local and small businesses is associated with low-obesity ones. Moreover, a curious peculiarity of Instagram is the use of hashtag \texttt{\#foodporn}. Rousseau~\cite{foodporn} discuss the use of this term and the potential dual-meaning of this connotation, where it can represent desirable and great food, or food which is unhealthy, which induces the feeling of ``guilt", and must be avoided. Below we further discuss the mixed signals in the community's approval that we discover in our dataset.


Using social media, we are able to observe the reaction of the other individuals to the posts, operationalized as \textit{likes} and comments. Some studies have focused on using social tagging of food pictures to encourage healthy eating~\cite{Linehan:2010:TST:1753846.1753980, Takeuchi:2014:USM:2638728.2641330}. For example, Stevenson~\etal~\cite{Adolescents} have interviewed 73 participants on perceptions of contradictory food-related social pressures and the negative self-perception generated by classifying adolescents preferred foods which may lead to a self-fulfilling pattern of unhealthy eating and notice challenges faces by individuals concerning the social pressures towards eating energy-dense foods on the one hand, and against obesity on the other. We second this finding by quantitatively showing the social interaction both healthy and unhealthy tags prompt around the pictures.

\section{Discussion}
\label{sec:discussions}

Data collected from the social media has well-known advantages and disadvantages (for a more involved discussion see \cite{abbar2014you}). In the case of Foursquare and Instagram, the bias is to more visited and shared locations, in comparison to the CHR data that is more comprehensive. There are several possible reasons, then, for the greater correlation of the share of fast food restaurants to obesity, compared to this ``offline'' resource. Our definition of fast food, or that of Foursquare users, may differ from that used by CHR. However, because of the ``popularity'' bias, it may also indicate a positive relationship between the popularity of fast food places and local obesity. To check the predictive power of Instagram data, we build a linear regression model to predict the obesity rate of the county in which a photo was taken, including restaurant, social approval, and hashtag features. This model, though showing significant coefficients, had an extremely low R$^2$ of 0.013, illustrating that without the deeper demographic knowledge, the picture metadata was not sufficient to predict county-wide obesity. 

A distinction -- not present in the previous studies -- we examine here, is one between local and chain restaurants, as empirically determined using the aggregate data. We find that users are almost twice as likely to post a picture at a local restaurant than a chain, and keep posting (which may indicate more frequent visits). We also find hashtags associated with small business like \texttt{\#eatlocal} and \texttt{\#smallbiz} to be associated with lower-obesity areas. The importance of these small restaurants is further underscored by the fact that, unlike large chains, it may be easier to work with them as a community to promote healthier menus, such as \emph{Eat Well! El Paso} program in Texas\footnote{\url{http://www.kvia.com/news/healthier-options-for-children-at-local-restaurants/26415336}} or \emph{Choose Health} in Los Angeles\footnote{\url{http://laist.com/2013/09/13/city_partners_with_local_restaurant.php}}. We also find that the chains which may not be considered as fast food could be popular for social gatherings, including \emph{The Cheesecake Factory}, \emph{Yard House}, and \emph{Dallas BBQ}. Thus, beyond fast food, such chains provide a social setting in which to share food, which may not be the healthiest choice. 

Our foray into the perception of food shows an alarming, overwhelming approval of addictive foods high in sugar and fat. Even though the users associate \texttt{\#foodporn} with potentially healthier cuisines (mainly Asian), the social approval of the images emphasizes the values social media reinforces. More work needs to be done to reveal the motivations behind users' interactions on social media, in order to harness them for the benefit of their dietary health.

\section{Conclusion}
\label{sec:conclusion}

In this paper we have taken a data-driven approach in analyzing food consumption on massive scale using Instagram and Foursquare. We used millions of posts from individuals in restaurants across the United States and correlated these with national statistics. Our analyses revealed a relationship between small businesses and local foods with obesity, with these restaurants getting more attention on these social media. However, the social approval, manifesting itself in the form of likes and comments, often favors the unhealthy dietary choices, favoring donuts, cupcakes, and other sweets.

In our future efforts we plan to use sentiment analysis techniques on picture comments in order to assess the level of approval and disapproval of the individuals' social network. We are also working on image recognition techniques using machine learning methods to enable inference of the content of the pictures and potential translation to caloric values. This will enable an increase in the confidence of our analysis for all locations and improve our ability to understand the actual food consumed.

\section*{Acknowledgments}
We appreciate support from Haewoon Kwak in data collection.

\balancecolumns
\end{document}